\documentclass[sn-mathphys-num,iicol]{sn-jnl}
\usepackage{graphicx}%
\usepackage{multirow}%
\usepackage{amsmath,amssymb,amsfonts}%
\usepackage{amsthm}%
\usepackage{mathrsfs}%
\usepackage[title]{appendix}%
\usepackage{xcolor}%
\usepackage{textcomp}%
\usepackage{verbatim}
\usepackage{subcaption}
\usepackage{manyfoot}%
\usepackage{booktabs}%
\usepackage{listings}%
\usepackage[utf8]{inputenc}
\usepackage{hyperref}
\usepackage{soul}

\theoremstyle{thmstyleone}%

\theoremstyle{thmstyletwo}%

\theoremstyle{thmstylethree}%

\begin{document}

\title[Article Title]{{Nightclub bar dynamics: statistics of serving times}}

\author{\fnm{Eduardo} \sur{V. Stock}}\email{eduardo.stock@ufrgs.br}
\author{\fnm{Roberto} \sur{da Silva}}\email{rdasilva@if.ufrgs.br}
\author{\fnm{Sebasti\'{a}n} \sur{Gon\c{c}alves}}\email{sgonc@if.ufrgs.br}

\affil{\orgdiv{Instituto de F\'{i}sica}, \orgname{Universidade Federal do Rio Grande do Sul}, \orgaddress{ \\ \city{Porto Alegre}, \state{Rio Grande do Sul}, \country{Brazil}}}

\abstract{In this work, we investigate the statistical properties of drink serving in a nightclub bar, utilizing a stochastic model to characterize pedestrian dynamics within the venue. Our model comprises a system of $n$ agents moving across an underlying square lattice of size $l$ representing the nightclub venue. Each agent can exist in one of three states: thirsty, served, or dancing. The dynamics governing the state changes are influenced by a memory time, denoted as $\tau$, which reflects their drinking habits. Agents' movement throughout the lattice is controlled by a parameter $\alpha$ which measures the impetus towards/away from the bar. We show that serving time distributions transition from a power-law to exponential and back to power-law as we increase $\alpha$ starting from a pure random walk scenario ($\alpha=0$). Specifically, when $\alpha=0$, a power-law distribution emerges due to the non-objectivity of the agents. As $\alpha$ moves into intermediate values, an exponential behavior is observed, as it becomes possible to mitigate the drastic jamming effects in this scenario. However, for higher $\alpha$ values, the power-law distribution resurfaces due to increased jamming. We also demonstrate that the average concentration of served, thirsty, and dancing agents provide a reliable indicator of when the system reaches a jammed state. Subsequently, we construct a comprehensive map of the system's stationary state, supporting the idea that for high densities, $\alpha$ is not relevant, but for lower densities, the optimal values of measurements occurs at high values of $\alpha$. To complete the analysis, we evaluate the conditional persistence, which measures the probability of an agent failing to receive their drink despite attempting to do so. In addition to contributing to the field of pedestrian dynamics, the present results serve as valuable indicators to assist commercial establishments in providing better services to their clients, tailored to the average drinking habits of their customers.}


\maketitle

\section{Introduction}\label{sec1}

The dynamics of pedestrian movement have long been of interest to researchers across various disciplines, ranging from physics\cite{shad2011} to urban planning\cite{baeza2021} and beyond. Understanding how individuals navigate through crowded spaces \cite{Schad2017}, such as city streets, public transportation systems, or entertainment venues, is crucial for optimizing infrastructures\cite{guo2023}, ensuring public safety\cite{ferenchak2024}, and enhancing overall efficiency\cite{wang2024}. 
In this context, physicists have been striving to identify the essential features of theoretical models required to replicate the significant phenomena arising from systems composed of self-driven agents.

With this in mind, a pioneering model was proposed by Helbing and Moln'{a}r~\cite{helbing-1995}, introducing the concept of a ``social'' force that directs crowds toward a common target, such as an exit or entrance door, particularly in scenarios of panic or stampedes. Other researchers have followed this concept,
seeking to understand the essential conditions for phenomena 
such as the formation of self organizing lanes~\cite{chraibi2015,mestrado-eduardo-roberto2017,li_2024} or clogging/jamming~\cite{nossoPRE2019,Stock-JSTAT-2019}, among others~\cite{stock2020,portz2010,silverber2013,oliveira2015,schad2018} to occur. At the same time, experimental observation of real events\cite{helbing2007,krausz2012,pastor2015,sticco2021,huarte2022} has been implemented in the attempt to map the intrinsic aspects of realistic crowd behavior.

A particular case of interest in pedestrian dynamics is the analysis of financial pressures within commercial venues and their potential correlation with an elevated risk of fatalities during tragic events. Such is the case of musical festivals or a nightclub party, for instance, where the unpreventable drug and acute alcohol consumption\cite{lojszczyk2023} has an important role in intentional and unintentional injuries\cite{clapp2017}.

In \cite{stock2023}, we explored a significant topic brought to light during the trial following the tragic \textit{Kiss Nightclub} fire\footnote{Brazil nightclub fire: Four convicted over blaze that killed 242,  \url{https://www.bbc.com/news/world-latin-america-59617508}}, which resulted in 242 casualties and left more than 600 injured~\cite{ponte2015,crestani2019}. During the event, a fire was sparked by the indoor use of firecrackers by the scheduled band, igniting the ceiling's soundproof layer. Prosecutors argued in the trial that the overcrowding of the nightclub, a fact confirmed by all witnesses including the defendants, was driven by a desire for maximum revenue and was the primary cause of the high casualty count.  In contrast, the defendants claimed that, even if the nightclub was at full capacity as it was, this situation was financially unfavourable once patrons were unable to access the bar impacting the revenue of drink sales.

The intricacies surrounding pedestrian dynamics within confined spaces such as bars or nightclubs, characterized by fluctuating densities, diverse movement strategies, and varying needs such as accessing drinks, pose a multifaceted challenge that demands comprehensive exploration. Whether driven by economic imperatives or concerns for the safety and comfort of patrons during events, this interdisciplinary issue warrants thorough investigation.

In this paper, our goal is to enhance the overall comprehension of pedestrian dynamics within service systems, particularly in bar environments. Through an examination of the statistical characteristics of serving times in a nightclub bar, we aim to uncover potential underlying patterns and behaviors that dictate the movement of individuals within these venues. Our analysis utilizes a stochastic model to depict the interactions between patrons and servers, offering insights into the dynamics of thirst, service, and movement within the bar setting.

Our research endeavors to offer more than just theoretical insights into pedestrian dynamics; we seek to provide practical guidance for designing and managing service systems in bustling environments. By delving into the fundamental principles governing pedestrian movement within bars, our aim is to improve the efficiency and overall enjoyment of experiences for both patrons and customers.

The contribution is organized as follows:

In the subsequent section, we introduce our stochastic agent-based model, detailing parameters and key variables such as serving times, concentration of states, and local conditional persistence. The latter quantifies the proportion of agents unable to obtain a drink when desiring one, drawing on concepts from coarsening literature \cite{Derrida}. We then proceed to present and analyze our findings. Finally, we summarize the results and draw conclusions.

\section{Model}\label{sec2}
\begin{figure*}[t!]
\centering
\includegraphics[width=1.0\textwidth]{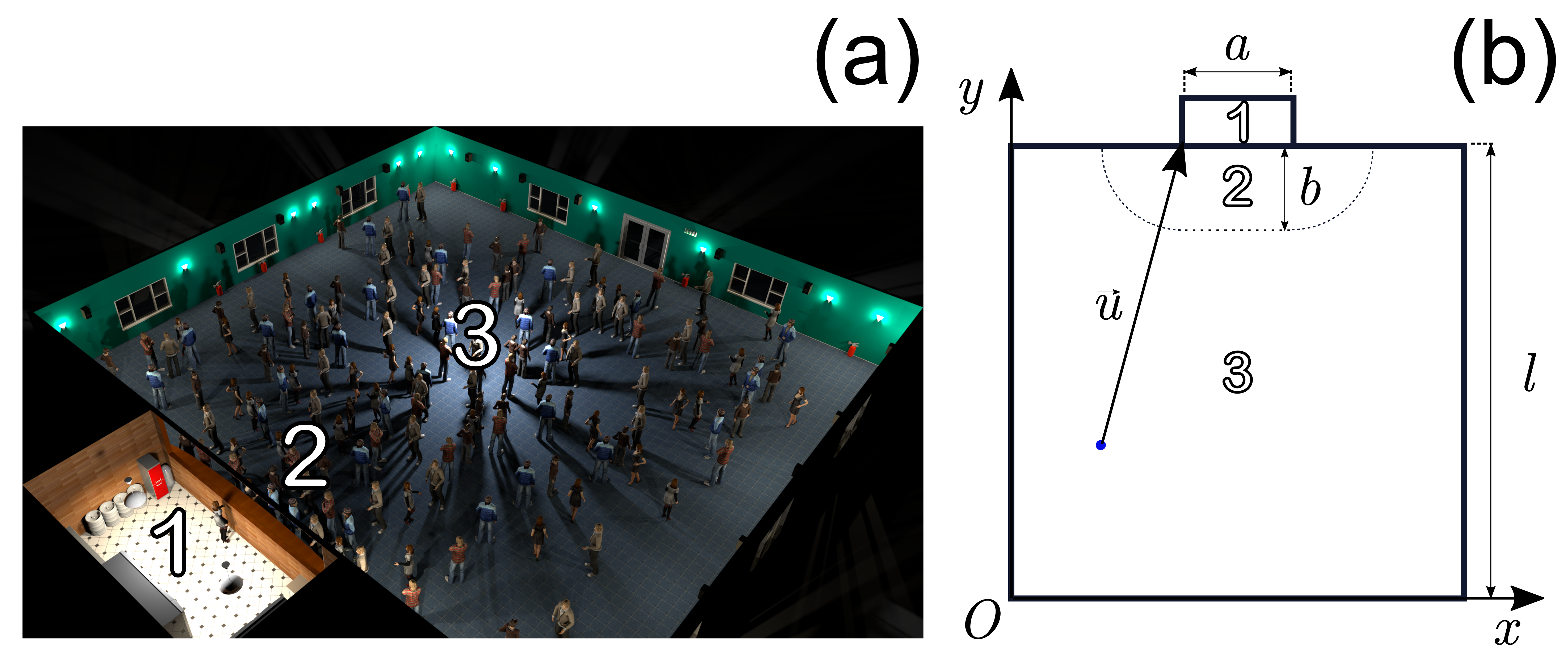}
\caption{(a) Visualization of a nightclub featuring a basic square layout with the bar situated along one side. (b) The lattice utilized in our model to replicate the nightclub comprises three distinct regions: (1) the bar, (2) the bar zone, and (3) the dance floor.}
\label{nightclub}
\end{figure*}

In this study, we adopt the model proposed by two of the authors \cite{stock2023}, which builds upon the lattice gas dynamics framework extensively explored in existing literature (e.g., see the framework by Katz et al.~\cite{katz1983}), adapting the transition probabilities initially outlined in \cite{roberto2015}. The model comprises a system of $n$ agents (patrons) capable of moving on an underlying square lattice with a side length of $l$, representing a nightclub. For simplicity, we designate the bar as a region outside the lattice (yet contiguous to it), centered on one of its sides with length $a$, thus dividing the nightclub venue into three primary regions: (1) the bar, (2) the bar zone, and (3) the nightclub dance floor, as illustrated in Fig.~\ref{nightclub}.
Region~(1), the bar, is inaccessible to the agents, who purchase drinks upon entering Region~(2), adjacent to the bar. 
The bar zone (Region (2)) is composed by all cells with distance no greater than $b$ from the bar and characterizes a turbulent region resulting from patrons attempting to reach the bar, others manoeuvring away from it with beverages, and agents moving randomly. 

During a party, each agent is in one of three possible states: (i) thirsty, (ii) served, and (iii) dancing; the number of agents in each state is denoted as $n_t$, $n_s$, and $n_d$, respectivelly, with the constraint $n_t+n_s+n_d=n$. As the dynamics evolves, agents change their states and the number of patrons within each states can be measured by its concentration in relation to the total number of agents $n$. Then, the concentration of agents in each of the three possible states are defined as $c_t\equiv n_t/n$,  $c_s\equiv n_s/n$, and  $c_d\equiv n_d/n$, with the constraint expressed as $c_t+c_s+c_d=1$.

Within the drinking dynamics, agents change their states in a cyclic fashion (i)$\rightarrow$(ii)$\rightarrow$(iii)$\rightarrow$(i) which comprehends the drinking dynamics as follows:
\begin{itemize}
\item An agent in state (i) will tend to move towards the bar to buy a drink. Once the agent reaches a cell contiguous to the bar, he/she immediately changes to state (ii);

\item once in state (ii), the agent will tend to move outside Region (2), changing to state (iii) immediately after leaving that region; 

\item this agent will remain in state (iii) for $\tau$ time steps until he/she wants another drink, so switching to state (i) again.  
\end{itemize}

Agents possess an individual characteristic drinking time, denoted as $\tau$, which represents the interval between drinks. Depending on the relationship between $\tau$ and the maximum duration of the party, denoted as $t_{max}$, an agent can be classified as alcoholic ($\tau \ll t_{max}$), abstemious ($\tau>t_{\max}$), or normal (with intermediate values of $\tau$).

With that in mind, between time steps $l$ and $l+1$, a given agent at cell $(i,j)$ can hop to anyone of its four neighboring cells, $(i',j')$, with the transition probability:
\begin{equation}
	Pr_{(i,j)\rightarrow {(i',j')}}^{(l)}=p+\alpha \left( \Delta \vec{r}\cdot
	\hat{u}\right) \sigma^{(l)},
	\label{probs}
\end{equation}%
where $p$ is the constant probability of making the transition --because of the isotropic possibility of choosing any of the four directions, $p=1/4$; $\alpha$ is the coefficient that regulates the degree of bias in the movement; $\Delta \vec{r}$ can be any of the four Cartesian movements
($\hat{e}{x}$, $\hat{e}{y}$, $-\hat{e}{x}$, and $-\hat{e}{y}$) with equal probability; $\hat{u}=\vec{u}/||\vec{u}||$ is the unit vector associated with the static floor field, which guides the agents toward the bar. $\sigma^{(l)}$ is the state of the agent associated with its drinking dynamics, previously referred, and that can assume the values
\begin{equation}
	\sigma^{(l)}=\left\{ 
	\begin{array}{ccc}
		1, & \text{if} & \text{state }(i), \\ 
		-1, & \text{if} & \text{state }(ii), \\ 
		0, & \text{if} & \text{state }(iii).%
	\end{array}%
	\right.  \label{func}
\end{equation}%

The inner product presented inside parenthesis on Eq.~\ref{probs} is the component of the displacement vector of the referred transition pointing in the static floor field direction, which reflects in an increment or decrement of the agent chances to move in that direction. The static floor field is responsible to guide patrons in and out of the bar, which happens whenever they are in states (i) or (ii). 
The factor $\alpha$ introduces asymmetry into the movement of patrons as they travel to and from the bar.
When considering transition probabilities, it's essential to adhere to the following constraint:
\begin{equation}
\begin{array}{ccc}
Pr_{(i,j)\rightarrow {(i,j)}}^{(l)}&=&1-\sum\limits_{\left\langle i\prime
,j\prime \right\rangle }Pr_{(i,j)\rightarrow {(i\prime ,j\prime )}%
}^{(l)}\\ &=&1-4p\text{,}
\end{array}
\end{equation}
this, along with our selection of $p=1/4$, implies that agents cannot remain stationary when they are chosen in the current MC-step and the cell to which they will transition is empty. The cells symbolized by $\left\langle i\prime
,j\prime \right\rangle$ denotes that the sum happens for the four first neighboring cells. 

Here, it is important again to clarify that $p=1/4$ does not imply that a particle necessarily remains stationary. Two scenarios illustrate when this occurs. Firstly, our computer simulations operate asynchronously; that is, we select a particle and determine its movement. If the selected destination is already occupied, the particle remains stationary. We repeat this process a number of times, exactly equal to the number of agents in the system. Some particles may be selected once, others with lesser probability, and in certain instances, some particles are not selected at all. In these cases, the particles also remain stationary. 

As a model to describe a nightclub dynamics, we opted to maintain the simplicity of the model, so it is important to clarify some points:
\begin{itemize}
\item Agents have an ideally infinity amount of money to buy their drinks within the time frame of a nightclub party;
\item the number of agents is fixed during the entirety of the party;
\item there is only one type of drink to be bought;
\item the drink is bought immediately after the agent gets to a bar cell, which means we are not considering any waiting time to be served due to drink preparation or queue related to it;
\item each agent's memory time $\tau$ is constant during the whole party;
\end{itemize}

Our focus in this work is to study the temporal features of the drinking dynamics and its relation with all system parameters. For that, we will first study the serving time distribution, denoted as $\Delta T$, which is the time an agent takes between wanting a drink and actually getting it. In practical terms, this random variable consists of the time interval that an agent spends in the state of thirstiness (state (i)). The serving time is one of the variables that influence the satisfaction felt by patrons within the nightclub experience.

Directly associated with the serving time is the patrons ability to actually get a drink given they wanted it. To measure this variable, we define the conditional local persistence of drinking, denoted as $f(t)$, which is the probability that a given agent do not drink for the first time until an instant $t$, given he/she was thirsty.

While the notion of persistence originated within the framework of spin system coarsening, specifically as local persistence (see, for example, \cite{Derrida, Silva2004}), its application has since broadened to encompass global parameters like magnetization rather than individual spins, facilitating the description of critical dynamics (see, for example, \cite{Majundar, Silva2003}). But the concept goes beyond with aplications in game theory, Econophysics, spatial exploration (see for example: \cite{Constantin,Silva2006,Silva2010,Silva2018}).  

We modify the concept of local persistence to account for the scenario in which agents desire a drink but do not obtain one. This adjustment results in a measure that does not exhibit monotonically decreasing behavior, contrary to the original definition. However, this deviation is not problematic for our purposes as we are going to explain in the subsequent section.

We implemented our model via Monte Carlo (MC) simulations considering a fixed number of agents $n$ throughout the party time frame and with $p=0.25$ for all cases. Also, we used in this work initial conditions considering agents positioned randomly throughout the lattice by using two independent uniform distributed random variables (\textit{ran2}). Each agent's initial biologic clock associated to their drinking habits was set randomly using \textit{ran2} with support $[0:\tau_i]$ for  $i=1,...,n$. This biologic clock counts the amount of time for the agent to reach the state of thirstiness.

Thus, in the next section we will show the main results of the statistics associated with time related variables that serves as reference of clients satisfaction. 

\section{Results}\label{sec3}

The first results that we show are related to the serving time distributions. Due to the random nature of data acquisition of serving times, we performed a non-fixed number of sequential experiments (open loop) with same parameters (but different seeds) to be sure that the serving time sample had a size of at least $n_{\mbox{sam}}=10^5$.

In the sequence, we also show the results associated with the conditional local persistence of drinking ($f(t)$). To measure this quantity, we calculate at each instant of the simulations the ratio between the number of thirsty agents that did not get any drink and the number of people that felt thirsty at least once. We defined this quantity in that way so it reflects only the inability to get a drink of the people that actually wanted it, and so no agent that was not thirsty was accounted for. As a consequence of its definition, the persistence can only be calculated after the first patron felt thirsty and so its value at each time step is averaged over a fixed number of identical experiments (same set of parameters, but with different seeds) but accounting only the ones that presented a value of $f(t)$. For the sake of persistence calculation, the number of experiments is denoted as $n_{\mbox{run}}$ and it is constrained to the relation $n \times n_{\mbox{run}}=10^6$.

\subsection{Serving time distributions}

We first studied the serving time distribution for the simplest case where there is no biased movement, i.e. $\alpha=0$ and $b=0$. We explored the influence of different bar sizes for a system of agents with same memory time ($\tau$) which can be achieved by a Dirac's Delta probability distribution $Pr(\tau)\propto \delta(\tau)$. We simulated the nightclub evolution for $t_{\max}=10^4$ MC steps on a lattice with side of length $l=64$ for a density of agents $\rho=0.25$.
\begin{figure}[h!]
\centering
\includegraphics[width=1.0\columnwidth]{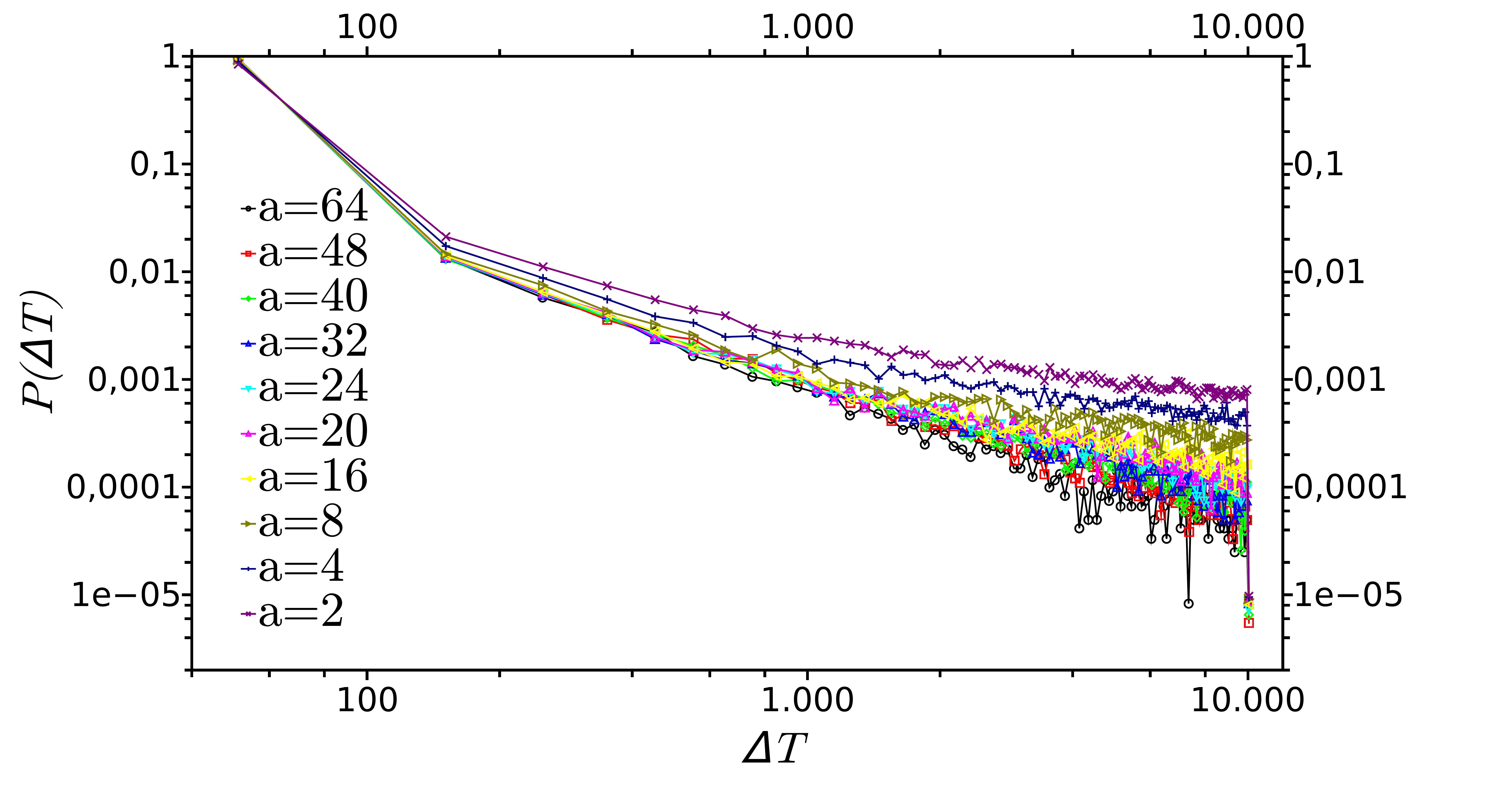}
\caption{ Distribution of serving times for different values of the bar length, $a$, for $l=64$, $b=0$, $\rho=0.25$, $\alpha=0$, and agents with memory time given by $P(\tau)\propto \delta(\tau)$. We note that serving times distribution have a power-law form, with its exponents decreasing slightly as $a$ increases.}
\label{fig:bar_size_study}
\end{figure}

In Fig. \ref{fig:bar_size_study}, we observe in a log-log scale that serving time distributions present a power-law shape but with slightly different exponents depending on the value of $a$. We note that as the size of the bar increases the exponent decreases, indicating that the unbiased motion of patrons results on a wide range of serving times, with people taking times up to the party time frame to get a drink.
\begin{figure}[h!]
\centering
\includegraphics[width=1.0\columnwidth]{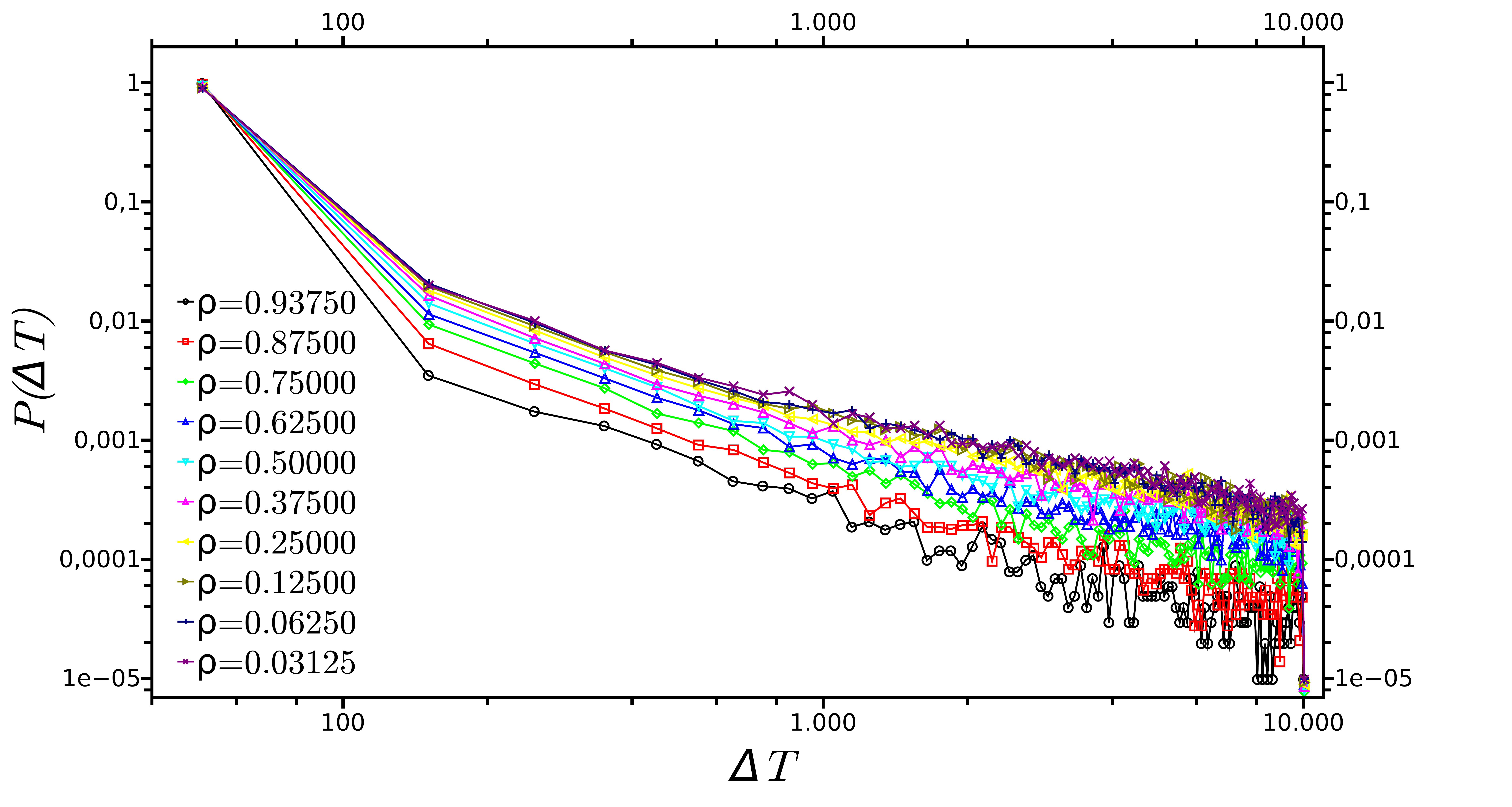}
\caption{ Distribution of serving times for different values of density for a system with $\alpha=0$, $a=16$, $b=0$, and memory times given by $P(\tau)\propto \delta(\tau)$. We can observe that the power-laws' exponent are not clearly influenced by the different values of $\rho$ for the most part of $\Delta T>100$ MC steps, even though a upward vertical shift (on log-log scale) is observed for increasing values of density. However, $\Delta T \le 100$ MC steps suggests that the power-law exponent decreases with density with a crossover value $\Delta T_c \approx 60$ MC steps for which $P(\Delta T_c)$ is the same regardless of $\rho$.}
\label{fig:density_study}
\end{figure}

Another important parameter of the dynamics is the density of patrons ($\rho$) which is kept fixed throughout the nightclub party. So, in Fig. \ref{fig:density_study}, we show the serving time distribution for different values of density for a system with same parameters as used in the Fig. \ref{fig:bar_size_study}, but with the difference that here we fixed $a=16$. We observe that the power-law exponent does not change with density for the most part of serving times. However, a larger number of people gets its drink rapidly for larger densities, which can be observed by the difference of inclination of the distribution curves for smaller values of $\Delta T$. This behaviour suggests a quasi-jamming scenario for higher density, where agents that are closer to the bar keep getting their drinks, even though they do not present a biased movement, while agents that are far from the bar rarely get their drinks.

\begin{figure}[th!]
\centering
\includegraphics[width=1.0\columnwidth]{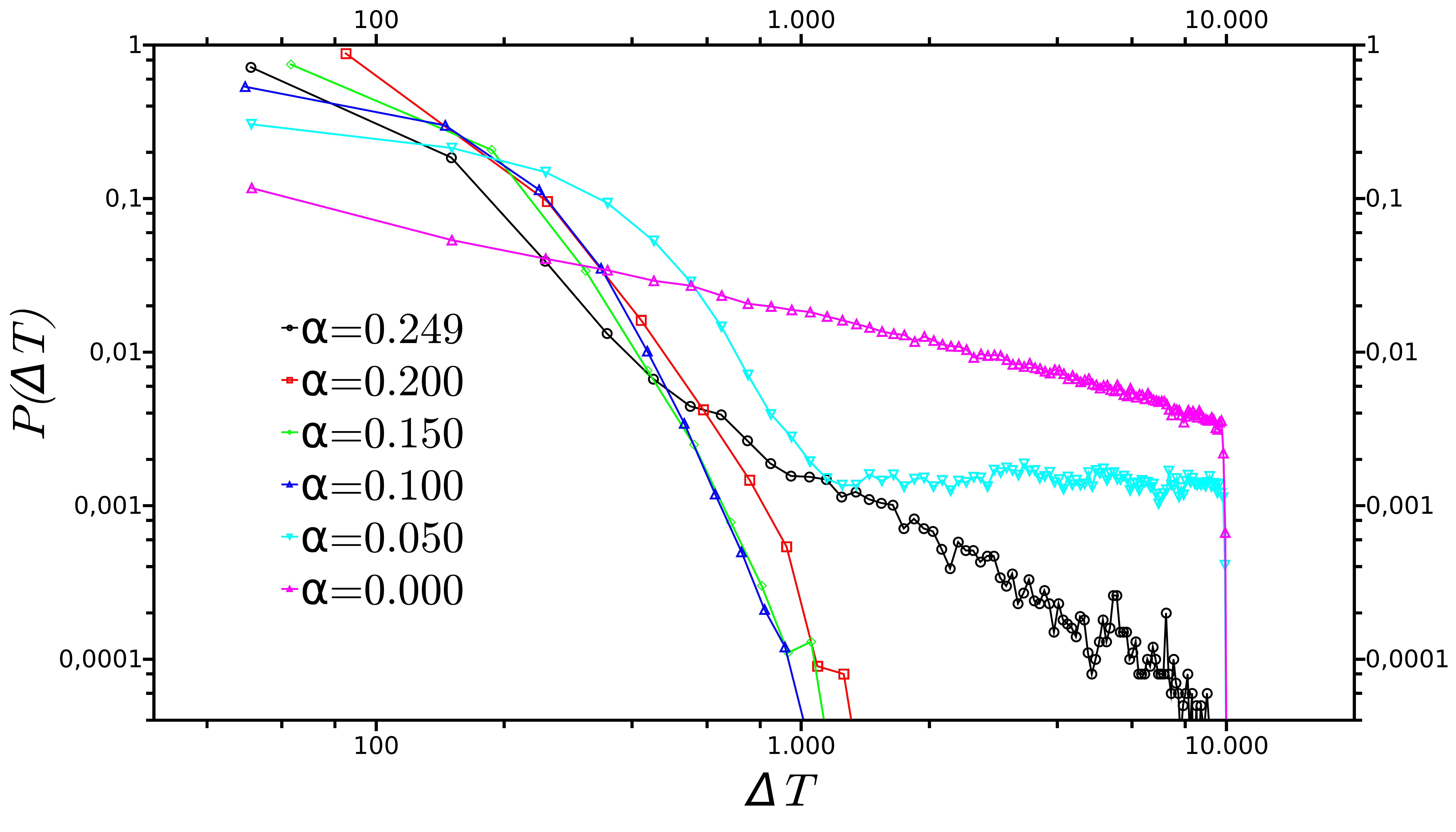}
\caption{ Distribution of serving times for different values of $\alpha$ for a system of side length $l=64$, bar dimension $a=16$, bar zone radius $b=12$, and a density of $\rho=0.25$ agents. In this study, we observe a transition in the serving time distribution during the party from a power-law form for $\alpha=0.0$, to an exponential behaviour $0.5 \le \alpha \le 0.15$, and a what appears to be a power-law distribution once again for $\alpha \approx 0.25$.}
\label{fig:alpha_study}
\end{figure}

Now we focus on the study of a more realist scenario, where patrons tend to move towards or away from the bar whenever they fell thirsty. 
In Fig. \ref{fig:alpha_study}, we show the serving time distribution for different values of $\alpha$, ranging from the previous studied  scenario of a lattice gas-like dynamics ($\alpha=0$) to a strongly driven dynamics ($\alpha \approx p$).

We notice that for $\alpha=0$ (Fig. \ref{fig:alpha_study}), the serving times follows the well defined power-law distribution, however a slightly increase on the patrons impetus as we see for $\alpha=0.05$ the serving time distribution start presenting two behaviours: an exponential distribution for serving times $\Delta T \lesssim 1000$ MC steps and uniform distribution for $\Delta T \geqslant 1000$ MC steps. This hybrid behaviour suggests that a non-trivial transition is occurring. As we study the interval $0.1 \lesssim \alpha \lesssim 0.2$, we note that the serving time distribution assumes a well defined exponential shape, but the support is shortened to approximately $[50,1000]$ MC steps suggesting that the level of efficiency of serving times is increased. However, by only studying the distribution of serving time we would not be able to differ a jamming scenario from an efficient serving service case. For instance, if jamming were to occur in a rather early time of the party in comparison with $t_{\max}$, we would not observe medium or long serving times, because theoretically the serving times would be infinite ($\Delta T \gg t_{\max}$). 

\subsection{Agents' state concentration}

\begin{figure}[th!]
\centering
\includegraphics[width=1.0\columnwidth]{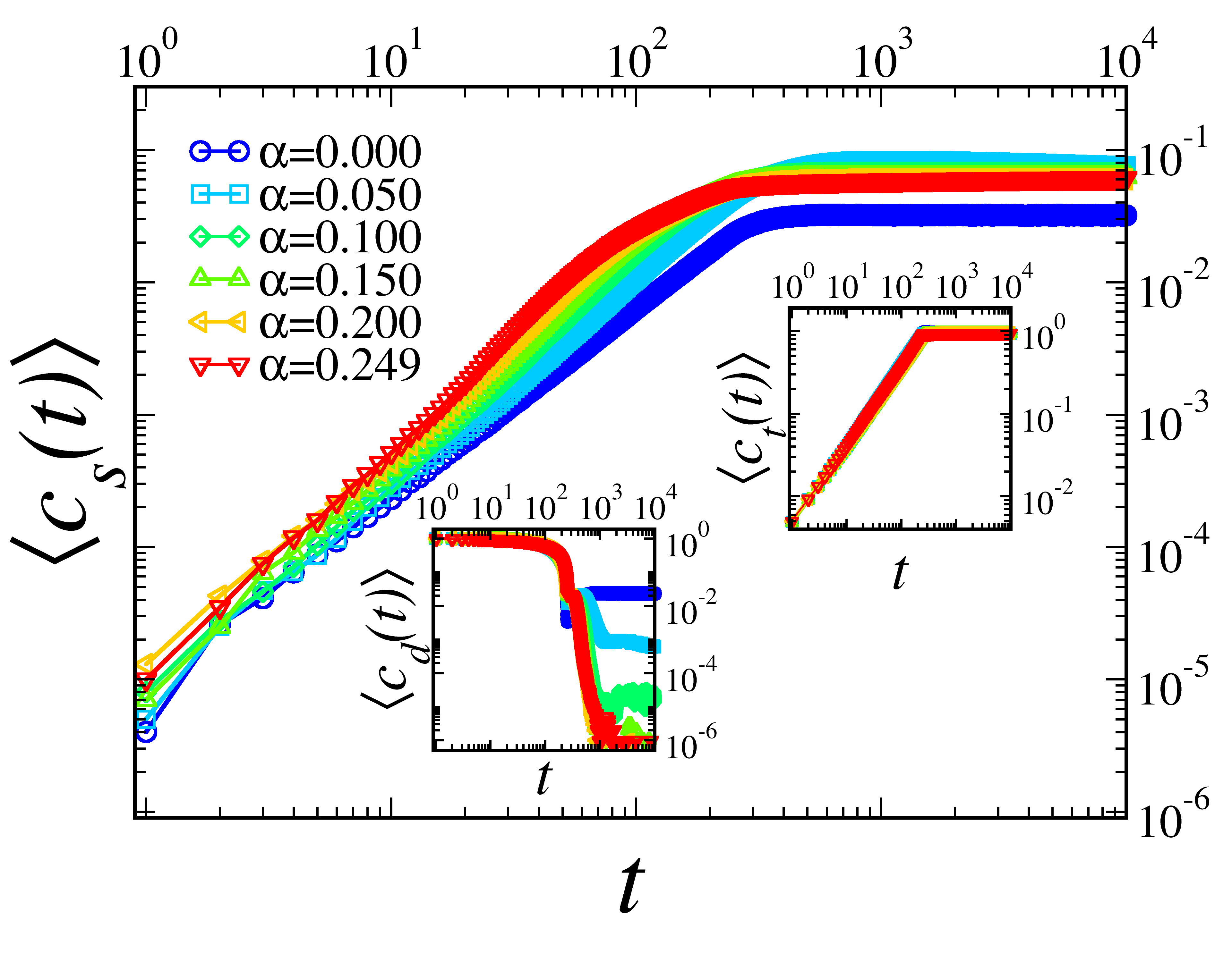}
\caption{Time series of the average concentration of served, thirsty, and dancing agents denoted as $\langle c_s(t)\rangle$,  $\langle c_t(t)\rangle$ (inset plot),  $\langle c_d(t)\rangle$ (inset plot) for different values of $\alpha$. As $\alpha$ increases, we identify the occurrence of jamming on the system once $\langle c_t(t)\rangle$ increases with time, at the same time that $\langle c_s(t)\rangle$ stagnates and $\langle c_d(t)\rangle$ drops to zero, which implies that people with drinks are stuck in the bar zone.}
\label{fig:concentration_study}
\end{figure}

To have a better understanding over the seemingly lower serving times for the scenario with biased patrons, we observe the evolution of the average concentration of served, thirsty, and dancing agents for the evolution of a system with same parameters used in Fig.\ref{fig:alpha_study}. In Fig. \ref{fig:concentration_study}, we show the evolution of $\langle c_s(t)\rangle$,  $\langle c_t(t)\rangle$ (inset plot),  $\langle c_d(t)\rangle$ (inset plot). We observe that for the simplest case ($\alpha=0$), the concentration of thirsty agents follow the trend of the other cases ($\alpha>0$), which is a consequence of the chosen memory time.

However, when observing the concentration of dancing and served agents in Fig. \ref{fig:concentration_study}, we note that is this simple case that has the greater number of dancing agents as well as the smaller number of served agents. This two variables accounts for the patrons that are able to leave the bar zone and make the state transition (ii)$\rightarrow$(iii). This transition is not facilitated when we observe greater values of $\alpha$. For $\alpha\ge0.05$, we note that the number of served agents increases, has a maximum, and fall again. For $\alpha=0.249$, corresponding to a regime of more eager thirsty patrons, we observe that served patrons get trapped more rapidly on the bar zone. This jamming also reflects on the number of dancing patrons, which drops to zero once the boundaries of regions (2) and (3) (where state (iii) agents are ``formed'') presents a gridlock.
 
To better understand the transition that happens when changing $\alpha$, we studied the average concentration of served patrons at the end of the party, i.e. $t=t_{\max}$, as a function of $\alpha$ and $\rho$. We studied both parameters $\alpha$ and $\rho$ within its possible ranges given by $0<\alpha<p$ and $0<\rho<1$ respectively.

As previous results suggests, the system dynamics is quite sensitivity with small increments of $\alpha$, thus we opted to increment each of its value geometrically, so that a greater number of small values of $\alpha$ could be studied in comparison with greater values of $\alpha$ within the proposed range. The algorithm used to implement such increment whilst performing trivial parallel simulations is shown in the Appendix \ref{secA1}, where we started from $\alpha_1=0.001$ to $\alpha_n=p$ divided into $n_{int}=50$ intervals and $m=10$ parallel simulations for the results presented in this work. Each point in the plane $(\alpha,\rho)$ was obtained by averaging $n_{\mbox{nrun}}=10^5$ samples with identical parameters, but different seeds. We implemented a simple arithmetic increment of the parameter $\rho$ diving the studied range into $n_{int}=50$ interval as well.

\begin{figure*}[t!]
\centering%
\begin{subfigure}[t]{0.32\textwidth}
\centering
\includegraphics[width=\textwidth,trim={1.5cm 0cm 0cm 0cm},clip]{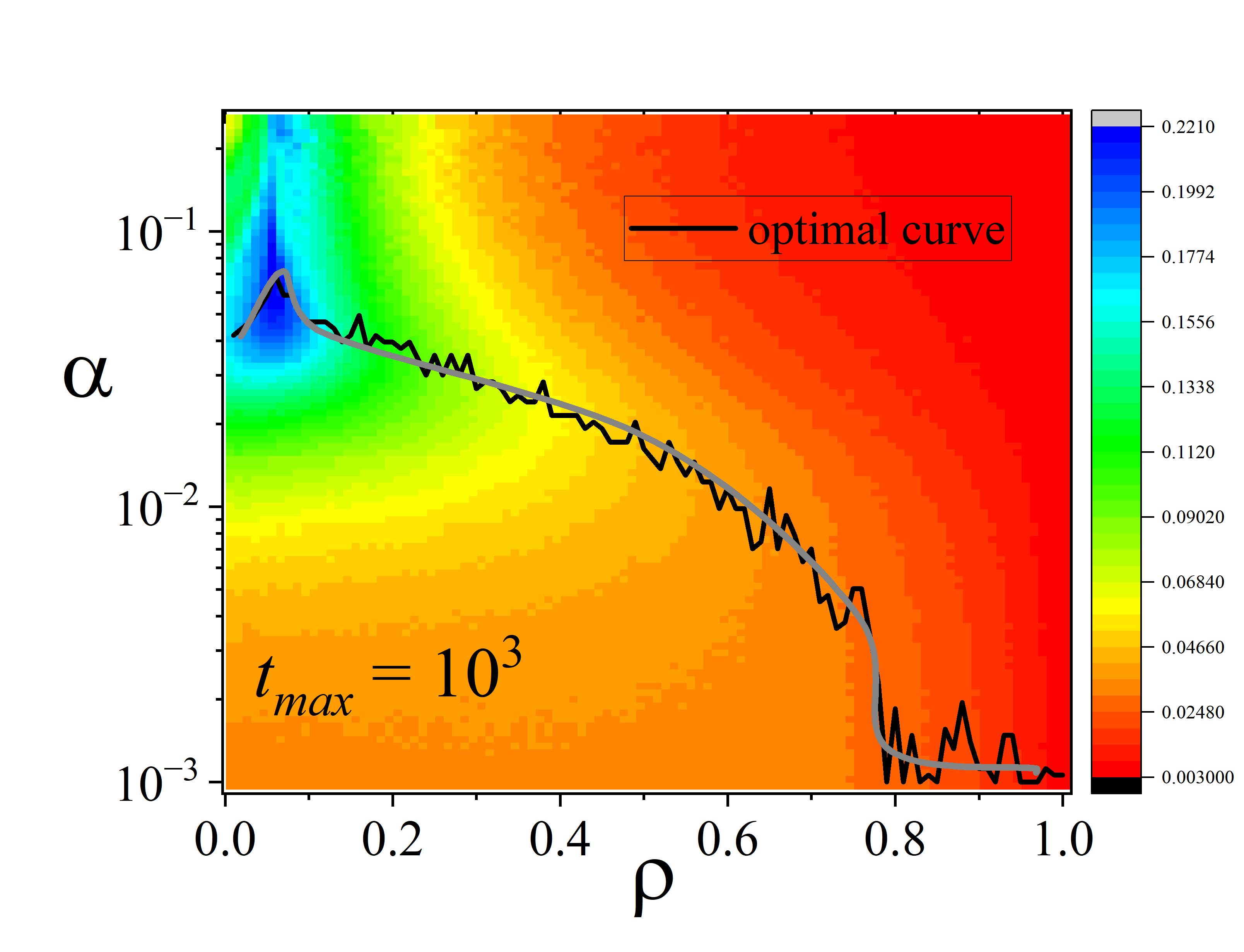} 
\caption{Heatmap of $\langle c_s \rangle$ for $t_{\max}=10^3$ with curve formed by maximum $\langle c_s \rangle (\alpha)$ (in black with curve of approximation in gray) for each value of $\rho$.} \label{fig:heatmap_1.1}
\end{subfigure}\hfill
\centering%
\begin{subfigure}[t]{0.32\textwidth}
\centering
\includegraphics[width=\textwidth,trim={1.5cm 0cm 0cm 0cm},clip]{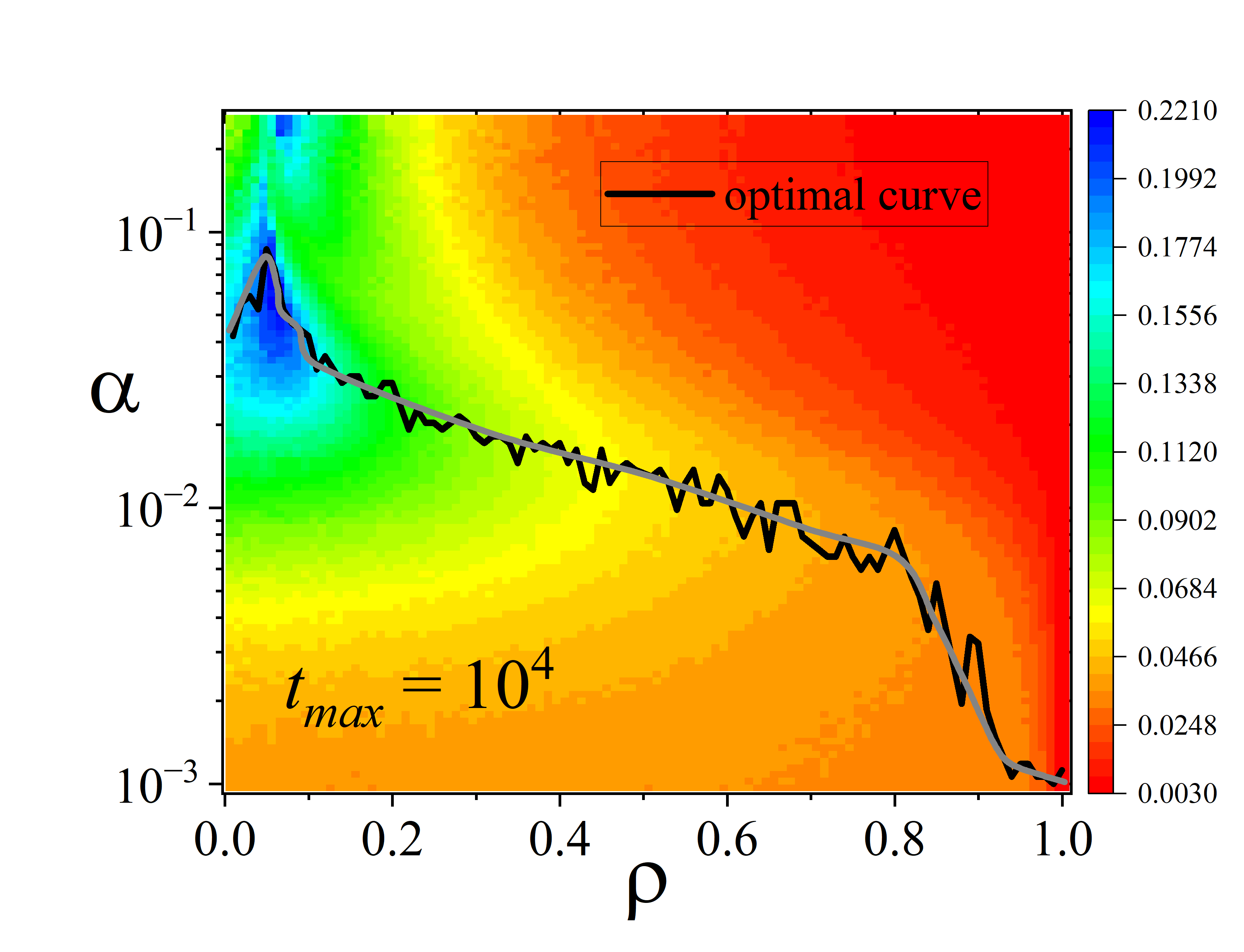} 
\caption{Heatmap of $\langle c_s \rangle$ for $t_{\max}=10^4$ with curve formed by maximum $\langle c_s \rangle (\alpha)$ (in black with curve of approximation in gray) for each value of $\rho$.} \label{fig:heatmap_1.2}
\end{subfigure}\hfill 
\begin{subfigure}[t]{0.32\textwidth}
\centering
\includegraphics[width=\textwidth,trim={1.5cm 0cm 0cm 0cm},clip]{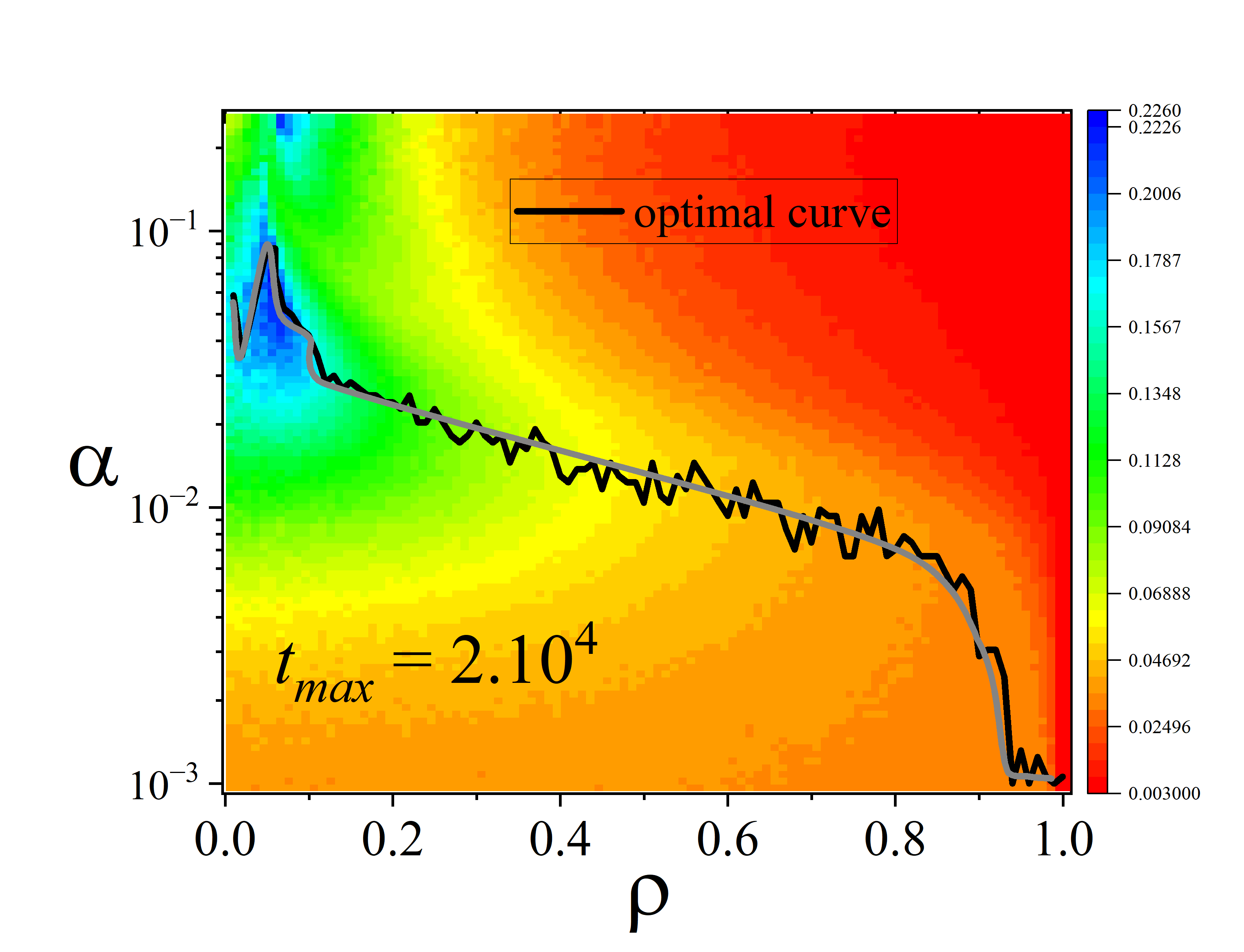} 
\caption{Heatmap of $\langle c_s \rangle$ for $t_{\max}=2\times10^4$ with curve formed by maximum $\langle c_s \rangle (\alpha)$ (in black with curve of approximation in gray) for each value of $\rho$.} \label{fig:heatmap_1.3}
\end{subfigure}
\caption{Heatmap showing the average density of served patrons at $t_{\max}=10000$ as a function of $\alpha$ and $\rho$. We note that the optimal curve of the average density of served patrons, $\langle c_s \rangle$, occurs for the optimal mobility for the largest densities.}
\label{fig:heatmaps_1}
\end{figure*}

In Fig. \ref{fig:heatmaps_1}, we show three color maps of $\langle c_s (t=t_{\max})\rangle$ {\em vs} $(\alpha,\rho)$ for $t_{\max}=10^3$ MC steps (a), $t_{\max}=10^4$ MC steps (b), and $t_{\max}=2 \times 10^4$ MC steps (c). For simplicity, we adopt the notations $\langle c_t (t=t_{\max})\rangle \equiv \langle c_t\rangle$,  $\langle c_s (t=t_{\max})\rangle \equiv \langle c_s\rangle$, and  $\langle c_d (t=t_{\max})\rangle \equiv \langle c_d\rangle$. We also show each corresponding optimized curve in black, which is formed by the set of points that maximize $\langle c_s \rangle$ {\em vs} $\alpha$ for each value of $\rho$. In gray, we show the approximated optimal curves. We note in each plot of Fig. \ref{fig:heatmaps_1} ((a), (b), and (c)) that a patterned behavior emerges where a maximum concentration of served agents happens for a specific region of low density and highly driven patrons observed in blue in Figs. \ref{fig:heatmap_1.1}, \ref{fig:heatmap_1.2}, and \ref{fig:heatmap_1.3}. This pattern does not change qualitatively within the three time frames studied, but solely looking at the served agents concentration is not sufficient to make a complete assessment on what actually is going on in the system. So we also studied the dancing and thirsty agents concentration.
\begin{figure*}[h!]
\centering%
\begin{subfigure}[t]{0.49\textwidth}
\centering
\includegraphics[width=\textwidth,trim={1.5cm 0cm 0cm 0cm},clip]{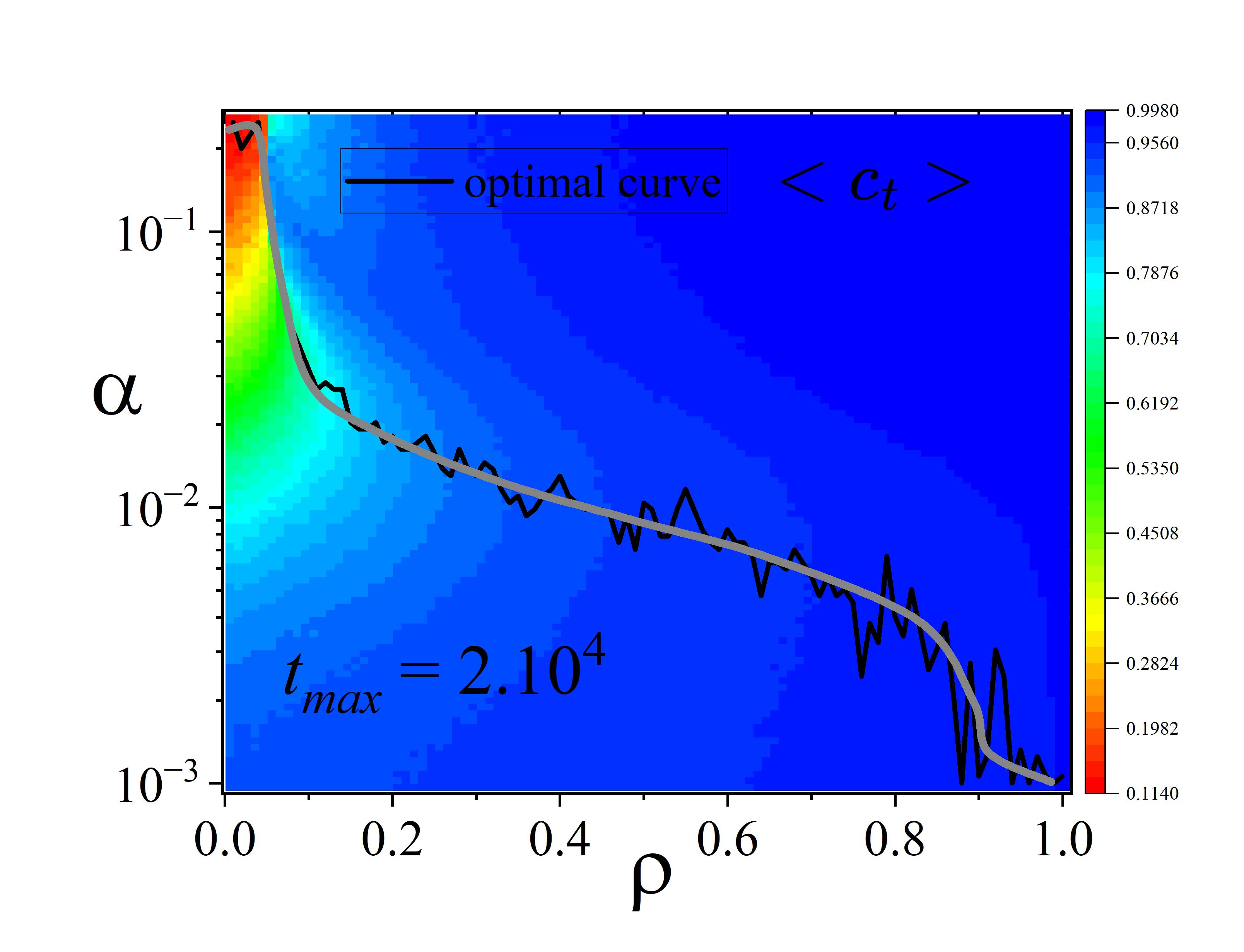} 
\caption{Heatmap of $\langle c_t(t_{\max})\rangle$ with curve formed by minimum $\langle c_t \rangle (\alpha)$ (in black with curve of approximation in gray) for each value of $\rho$.} \label{fig:heatmap_2.1}
\end{subfigure}\hfill
\begin{subfigure}[t]{0.49\textwidth}
\centering
\includegraphics[width=\textwidth,trim={1.5cm 0cm 0cm 0cm},clip]{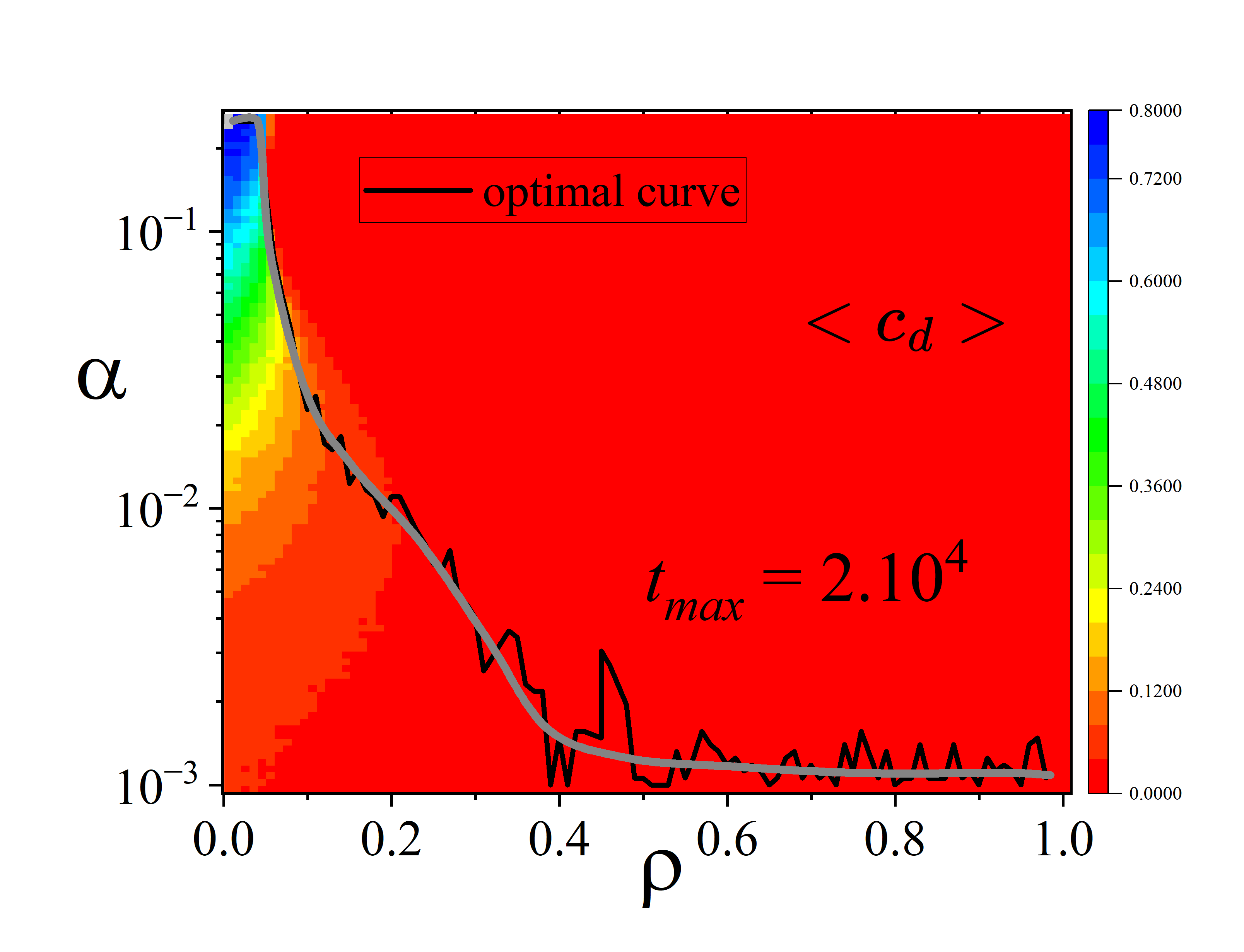} 
\caption{Heatmap of $\langle c_d(t_{\max})\rangle$ with curve formed by maximum $\langle c_d \rangle (\alpha)$ (in black with curve of approximation in gray) for each value of $\rho$.} \label{fig:heatmap_2.2}
\end{subfigure}
\caption{Heatmap showing the average concentration of thirsty patrons (a) and dancing patrons (b) as a function of $\alpha$ and $\rho$ as well as its respective optimal curves in black and its optimal approximation curve in gray. We observe that the region of minimum (maximum) concentration of thirsty (dancing) agents occurs for a low density, at approximately $\rho \leq 0.1$, of highly driven patrons, at approximately $0.1 \leq \alpha \leq 0.249$ , which corresponds to the optimal serving service.}
\label{fig:heatmaps_2}
\end{figure*}

In Fig. \ref{fig:heatmaps_2}, we show the color maps of the average concentration of thirsty and dancing agents for $t_{\max}=2\times 10^4$ obtained by using the same set of parameters in Fig. \ref{fig:heatmaps_1}. We also show in each plot the corresponding optimal curves in black and its approximation in gray, whereas the optimal curve corresponds to the set of points that minimize (maximize) $\langle c_d\rangle(\alpha)$ ($\langle c_t\rangle(\alpha)$) for each values of $\rho$ in Fig. \ref{fig:heatmap_2.1} (\ref{fig:heatmap_2.2}). We observe that the concentration of thirsty and served agents have similar optimal curves, which results from the fact that both concentrations are bound by a jamming scenario. As a matter of fact, the concentration of dancing agents can be even more reliable in terms of identifying a jamming occurrence, once its population drops to nearly zero in every case that the access to the bar is compromised. So we are able to identify that the optimal flow happens when the concentration of dancing agents is maximum, which we identify to occur for a low density of highly driven patrons as shown by the blue region in Fig. \ref{fig:heatmap_2.2}.

\subsection{Local Conditional Persistence}

\begin{figure}[h]
\centering
\includegraphics[width=1.0\columnwidth]{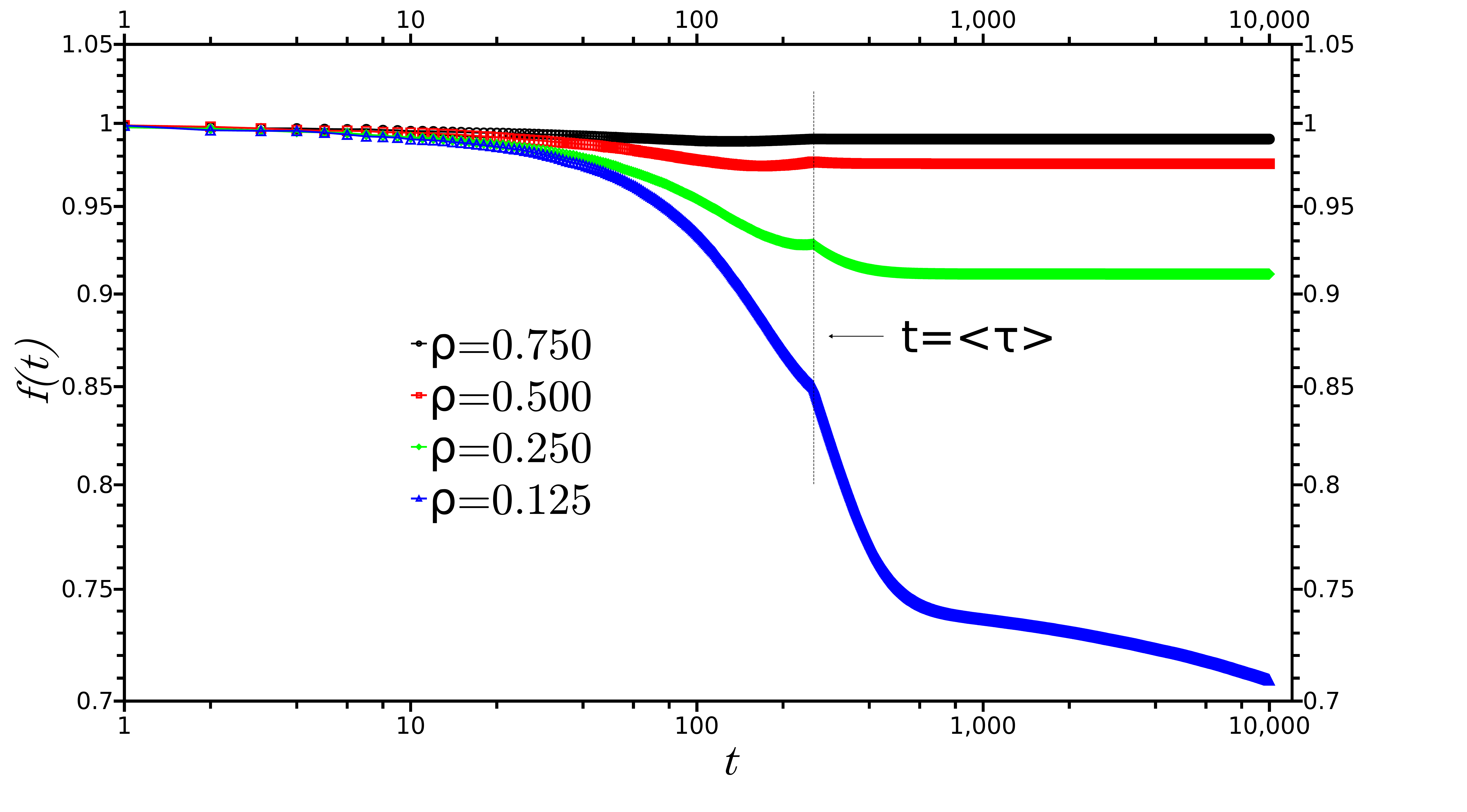}
\caption{ Conditional persistence time series for different values of $\rho$ considering a system with $l=64$, $\alpha=0.1$, $a=16$, $b=12$, and $P(\tau)\propto \delta(\tau-256)$ to define agent`s memory time. We note that a fairly empty nightclub, $\rho>0.125$, guarantees that around $30\%$ of consumers satisfied with serving service in comparison with the a nightclub filled up to $\rho=0.25$.}
\label{fig:persistence_density_study}
\end{figure}

As another important variable to measure the overall clientele's satisfaction and even identify jamming occurrence is the conditional local persistence ($f(t)$) defined at the end of the previous section. In the Fig. \ref{fig:persistence_density_study}, we studied the effects of different densities of agents on the persistence for a system with $l=64$, $\alpha=0.1$, $a=16$, $b=12$, and $P(\tau)\propto \delta(\tau-256)$. The dashed vertical line identifies the transient time interval, $t\le \langle \tau \rangle$, where the number of thirsty agents is still growing and after which every agent has felt thirsty at least once. This transient interval is easily known if Dirac's Delta are chosen to define agent's memory time. We observe that for $\rho\ge 0.25$ the persistence presents a stagnation, which is resulting from the jamming that we already knew happened for that particular choice of $\alpha$ and $\rho$ values. This persistence stagnation is also observed on in systems of spins \cite{Silva2004}. We also note that for $\rho=0.125$ the persistence drops monotonically for $t\ge \langle \tau \rangle$ suggesting that this particular density lies near the transition from a flowing to a jammed regime.

\begin{figure}[h]
\centering
\includegraphics[width=1.0\columnwidth]{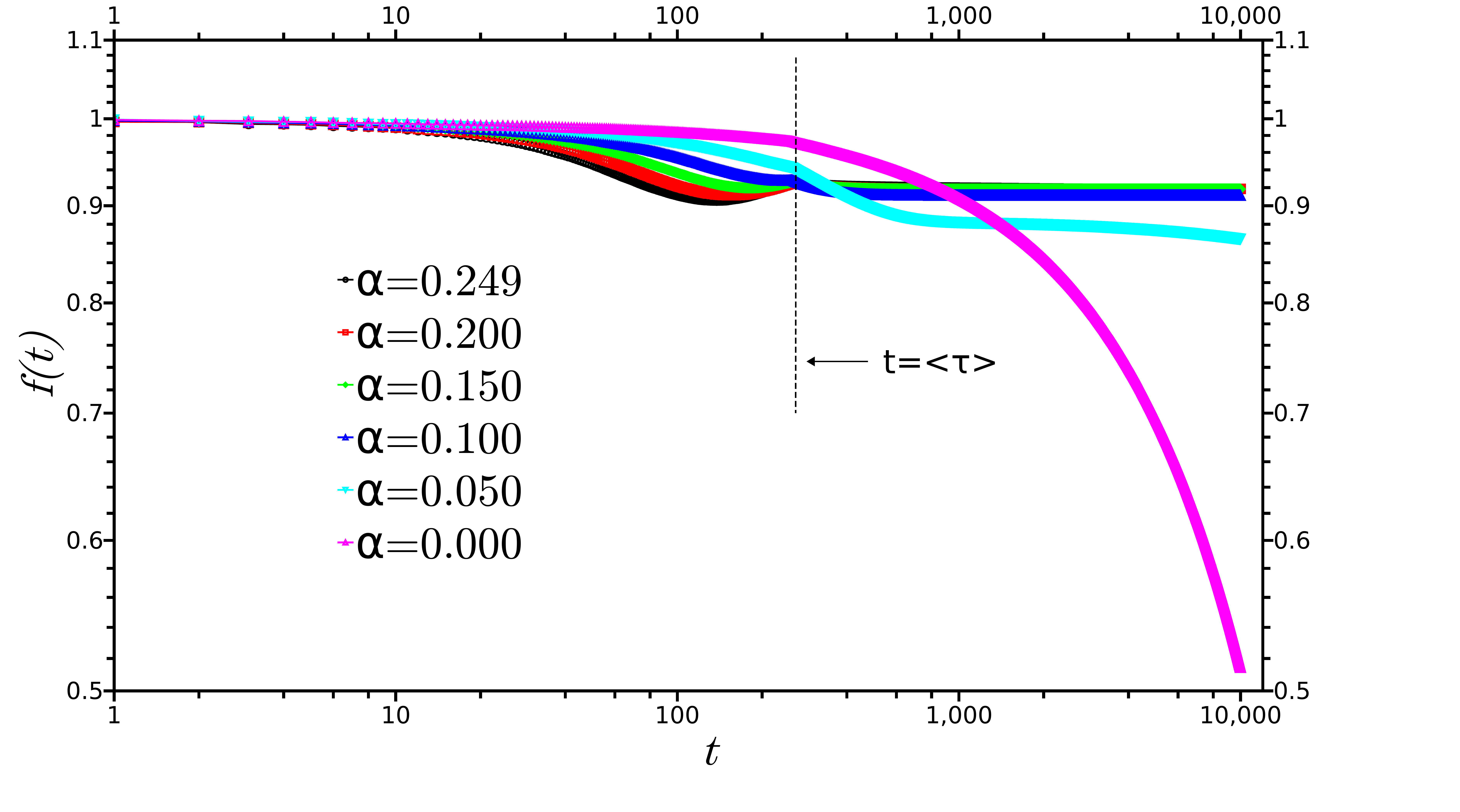}
\caption{ Conditional persistence time series for different values of $\alpha$ considering a system with $l=64$, $\rho=0.25$, $a=16$, $b=12$, and $P(\tau)\propto \delta(\tau-256)$ to define agent`s memory time. We observe that for the considered density, a slight increase of $\alpha$ relative to the standard lattice gas regime is sufficient to make more than $85\%$ of patrons unable to get a drink within the party time frame.}
\label{fig:persistence_alpha_study}
\end{figure}

We looked at the influence of different $\alpha$ on the persistence for a system with identical parameters as previously used except for the density $\rho=0.25$. In Fig. \ref{fig:persistence_alpha_study}, we note that after the transient interval ($t\le \tau$) for $\alpha\ge 0.1$ the system jams at that density, whilst for $\alpha=0.05$ the persistence drops monotonically with time. The most interesting result, though, is that the non-biased case appears to be the best scenario because almost half the patrons are able to drink within the party time frame studied. This observation comes as a result of a limitation of the lattice gas model, where hard body exclusion do not allow agents to pass through crowded areas, thus any minor bias towards/away from the bar is sufficient to lead the system  to a gridlock state. 

\begin{figure}[h]
\centering
\includegraphics[width=1.0\columnwidth]{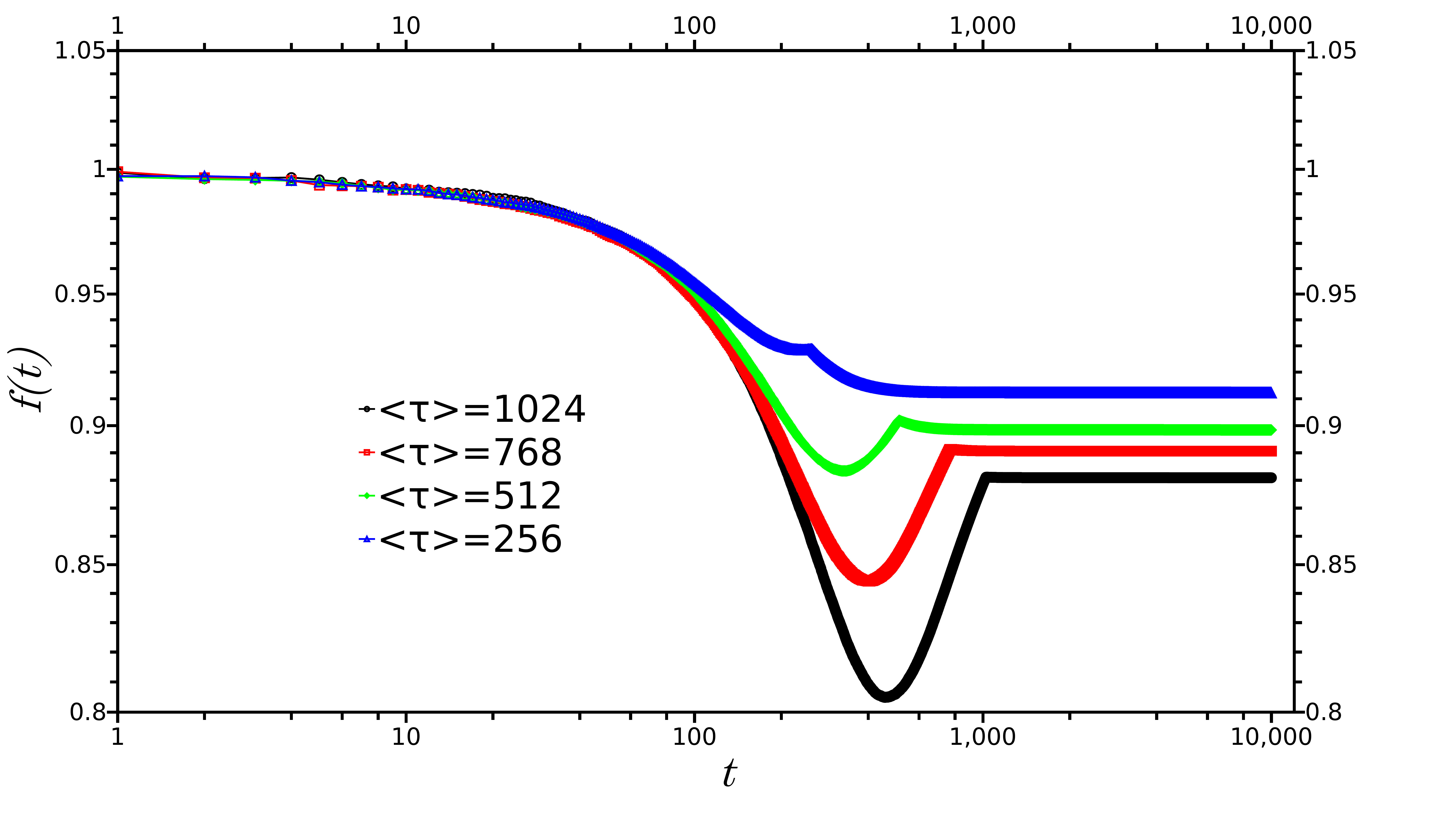}
\caption{Conditional persistence time series for different memory times considering a system with $l=64$, $\rho=0.25$, $\alpha=0.1$ $a=16$, and $b=12$. We note that the fraction of patrons that manage to get a drink increases with the $\langle \tau \rangle$ once the transient time ($t\le \langle \tau \rangle$) is larger, which results in a lower number of people trying to reach or leave the bar.}
\label{fig:persistence_memory_study}
\end{figure}

Finally, in Fig. \ref{fig:persistence_memory_study}, we show the influence of different average memory times on the persistence for a system with $\rho=0.25$, $\alpha=0.1$, $a=16$, and $b=12$. We note that increasing $\langle \tau \rangle$ immediately enlarges the transient time interval, $t\le\langle \tau \rangle$, which allows that a greater fraction of agents are able to drink within the party time frame, even though the system itself evolves to a jammed state. Even if in each case agents have the same memory time, their biologic clock is not synchronized duo to the initial conditions and the resulting larger transient time interval reflects on a delayed jammed scenario.

\section{Conclusion}\label{sec4}

In this contribution, we investigated key features in the dynamics of
patrons (pedestrian) in a nightclub, specifically regarding the process of acquiring drinks.
We focus on a traditional topology that encompasses dancing and consumption typical of nightclub environments. Our model introduces memory to drinking behavior, guiding patrons to the bar when they seek service. Through these dynamics, we unveil intriguing findings that support the notion that strong guidance is beneficial only up to a critical density of agents/patrons. At higher densities, approximately $\rho \approx 1/4$, we find that exponential serving times emerge only for intermediate $\alpha$ values.

Jammed conditions lead to a high probability of patrons failing to obtain drinks. We depict this phenomenon using heat maps, illustrating, as functions of $\rho$ and $\alpha$, the average density of served patrons, the average concentration of thirsty patrons, and dancing patrons. Additionally, we present optimal curves overlaid on these maps, showcasing the ideal conditions for drink acquisition based on $\rho$ and $\alpha$.

Moreover, we extend the concept of local persistence, which originated from coarsening dynamics, to quantify the probability of a patron not being served up to time $t$, given their desire to drink. This conditional form of persistence capture the monotonic decrease observed in the original concept only after the transient time ($t\le \tau$) where the number of agents getting thirsty is still increasing has passed. Moreover, the monotonic decay occurs for a system of biased agents ($\alpha>0$) but only for densities low enough so that jamming does not occur. Also, stagnation of persistence for when the system relax to a jammed state is also observed in our results. Despite this deviation caused by the transient time until all agents present the urge to drink, i.e. persistence conditioned to only thirsty agents, our adaptation aligns with the original concept, as evidenced by the heat maps, indicating that persistence remains constant over time for certain $\alpha$ values, akin to observations in coarsening dynamics.

Our work represents an interesting way of approaching this complex system of agents that comprehends the merge of commerce and pedestrian dynamics. It also can indicates statistical insights on the serving services in enclosed environments whilst considering profit of the nightclub owners and overall clientele's satisfaction. Our model, however, can be further updated by including even more realistic aspects in regards to the agents' drinking dynamics as well the possibility to include a viscosity-like feature, for instance, to overcome some limitations of the lattice gas model as addressed in this work\cite{scramble2023}.      

\backmatter

\bmhead{Acknowledgments}

R. da Silva would like to thank the CNPq for partly support this work, under the grant: 304575/2022-4. Eduardo V. Stock also would like thank CNPq for partly supporting this work under grant number 152715/2022-3.

\begin{appendices}

\section{Trivial parallelization algorithm}\label{secA1}

To better investigate the transition of the system that happens for small values of $\alpha$, we implemented its increment geometrically. In that sense, if $\alpha_1$ is the initial value used in the simulation and we want to observe the system' behaviour for $n$ different values of $\alpha$ ranging from $0 < \alpha \leq p$, we increment $\alpha$ by a fraction $\Delta \alpha$ of each previous value, then the $n$-th value of $\alpha$ will be
\begin{equation}
    \alpha_{n}=\alpha_1(1+\Delta \alpha)^{n-1}.
    \label{eq1}
\end{equation}

Now, to parallelized our code trivially, we must now divide the interval within $\alpha$ in $m$ parts where $n$ is a multiple of $m$, for simplicity, i.e., $n=\pi m$, where $\pi$ is a positive integer representing the number of values of $\alpha$ that each of our $m$ codes will observe. We will then, have
\begin{equation}
    \begin{array}{ccc}
    &\underbrace{\alpha_1 < ... < \alpha_{\pi}}_{\mbox{first code}}& < \underbrace{ \alpha_{\pi+1} < ... < \alpha_{2\pi}}_{\mbox{second code}} < \\ \\ &...& < \underbrace{\alpha_{n-\pi+1} < ... < \alpha_{n}}_{\mbox{m-th code}},
    \end{array}
\end{equation}
and the relation between the first and last values of $\alpha$ to be studied at the $j$-th interval, where $j=1,...,m$, is
\begin{equation}
\alpha_{j\pi} = \alpha_{(j-1)\pi+1}(1+\Delta \alpha)^{\pi-1} = \alpha_{1}(1+\Delta \alpha)^{j\pi-1},
\end{equation}
where the increment factor $\Delta \alpha$ is constant defined by the first and last values of $\alpha$ we want to study as given by Eq.\ref{eq1}:
\begin{equation}
    \Delta \alpha = \left(\frac{\alpha_{n}}{\alpha_1}\right)^{\frac{1}{n-1}}-1.
    \label{increment}
\end{equation}

Within a certain interval $j$, the $k$-th value of $\alpha$, with $k=1,...,\pi$, is determined by
\begin{equation}
\begin{array}{ccc}
 \alpha_{(j-1)\pi+k} & =& \alpha_{(j-1)\pi+k-1}(1+\Delta \alpha)\\ \\ &= &\alpha_{(j-1)\pi}(1+\Delta \alpha)^{k-1}\\ \\
 &= &\alpha_{1}(1+\Delta \alpha)^{k-1+(j-1)\pi}
\end{array}
\end{equation}
Thus, the parameters necessary to implement the trivial parallelization, where each code is design to study a different interval of $\alpha$, are
\begin{itemize}
    \item n: number of values of $\alpha$ to be studied in total;
    \item m: number of codes to run in parallel (must be chosen such that $n=\pi m$ with $\pi$ positive integer);
    \item $\alpha_1$: initial value of $\alpha$;
    \item $\alpha_n$: final value of $\alpha$;
\end{itemize}

\subsection*{Example of application}

\hspace{1cm}If we chose to study $n=50$ values of $\alpha$ within the range $0<\alpha \leq p$, we can have $\alpha_1=0.02$ and $\alpha_{50}=p$. We then chose to have $m=5$ codes running trivially in parallel (could be more depending on the urgency!), so that $\pi=10$ different values of $\alpha$ are studied in each interval.

The $j$-th code, will then have initial and final values of $\alpha$, respectively, 
\begin{equation}
    \begin{array}{ccc}
         \alpha_{(j-1)\pi +1} &=& \alpha_{1}(1+\Delta \alpha)^{(j-1)\pi} \mbox{\quad and}\\ \\
         \alpha_{j\pi} &=& \alpha_{1}(1+\Delta \alpha)^{j\pi-1}
    \end{array}
\end{equation}
where $\Delta \alpha$ is given by Eq. \ref{increment}.

\end{appendices}

\section*{Declarations}

\bmhead{Author contribution}

All three authors contributed equally to this work. They jointly conceived and designed the analysis, conducted formal analyses, wrote the paper, elaborated on the algorithms, analyzed the results, and reviewed the manuscript.

\bmhead{Data availability}

The data supporting the findings of this study are available upon request. Interested parties can access the data by contacting the authors via the email addresses provided in the contact information at the beginning of the article.

\end{document}